\documentclass[aps,onecolumn,floats,floatfix,showpacs,superscriptaddress]{revtex4}
\usepackage{graphicx,amssymb,amsmath}
\usepackage[latin1]{inputenc}
\usepackage{color}

\begin{document}

\title{Exploring the complex dynamics of a Duffing oscillator with ordinal patterns analysis}

\author{Maximillian Trostel}
\affiliation{Department of Physics and Astronomy, Carleton College, Northfield, 55057, USA}
\author{Mosses Misplon}
\affiliation{Department of Physics and Astronomy, Carleton College, Northfield, 55057, USA}
\author{Andr\'{e}s Aragoneses}
\affiliation{Department of Physics and Astronomy, Carleton College, Northfield, 55057, USA}
\affiliation{andres.aragoneses@planetadavinci.com}
\author{Arjendu Pattanajak}
\affiliation{Department of Physics and Astronomy, Carleton College, Northfield, 55057, USA}

\begin{abstract}

The driven double-well Duffing oscillator is a well-studied system that manifests a wide variety of dynamics, from periodic behavior to chaos, and describing a diverse array of physical systems. It has been shown to be relevant in understanding chaos in the classical to quantum transition. Here we explore the complexity of its dynamics in the classical and semi-classical regimes, using the technique of ordinal pattern analysis. This is of particular relevance to potential experiments in the semi-classical regime. We unveil different dynamical regimes within the chaotic range, which cannot be detected with more traditional statistical tools. These regimes are characterized by different hierarchies and probabilities of the ordinal patterns. Correlation between the Lyapunov exponent and the permutation entropy is revealed that leads to interpret dips in the Lyapunov exponent as transitions in the dynamics of the system.

\end{abstract}

\maketitle
 

\section{Introduction.}

The question of how classical chaos arises from underlying quantum mechanisms has long been a question of deep fundamental interest. Since closed quantum systems are periodic due to their discrete spectrum it is unclear how chaos (i.e. exponential sensitivity to initial conditions) emerges in the classical limit. Pioneering theoretical work [1,2] showed that quantum chaotic Poincare sections indicating chaos do in fact arise in open quantum systems (which limit to dissipative classical dynamics). 
Recent work has extended this to show that this chaos has a smooth quantitative quantum-classical transition [3]. This has been established using the Lyapunov exponent, which is the inverse of the time rate of exponential growth of small differences between two initial conditions (and hence is a measure of the dynamical complexity [4]). This transition also contains complicated non-monotonic behavior as a function of system size [3] for a particular paradigmatic system, the Duffing oscillator. That is, the results show that Duffing oscillators in the semi-classical regime (intermediate size) can display chaos which does not exist in either the classical (large size limit) or quantum (very small size limit) systems. The nature of the transition between the quantum and classical limit seems to be strongly dependent on the dissipation (coupling to the environment) for the system, and in particular on how that dissipation changes the details of the behavior of the classical limit of the system [3], for example whether the dynamics have high order periodic orbits or not.

While the Lyapunov exponent is a definitive measure of chaos and dynamical complexity and relatively straightforward to compute theoretically, this does not translate to experiments because Lyapunov exponents depend on details of phase-space trajectories which are not easy to access experimentally. Even with access to phase-space trajectories, the formal definition requires the ability to start two phase-space trajectories close to each other, and to be able to reset this evolution for one of the trajectories at intervals defined by the evolution itself, while forcing both trajectories to evolve with the same `noise' (due to environmental fluctuations). While experiments rarely allow the ability to control all these aspects of trajectories, they usually allow access to time-series for a single variable. This constraint is particularly of relevance in comparing classical and semi-classical dynamics because the size of the effective phase-space changes in the transition to quantum mechanics (as we discuss in more detail below) and the extremely complicated nature of incorporating quantum measurement and initial condition specification into any quantum system dynamics. 

To address these issues, in this paper we present results from using the technique of 'ordinal pattern analysis' and permutation entropy for the same semi-classical and classical Duffing oscillator systems as previously studied. We show that this allows us to distinguish quantitatively between different kinds of complexity using only time series data. Thus, this form of analysis complements Lyapunov exponent analysis and is of particular value in understanding quantum dynamical chaos and potential experiments [5] searching for such behavior.

There is a further benefit to using ordinal pattern analysis. Lyapunov exponents are a blunt measure of chaotic dynamics. That is, it is possible to have the same Lyapunov exponent and significantly different chaotic attractors and underlying dynamics. It is similarly possible to have very different Lyapunov exponents but have essentially the same dynamics topologically. This similarity and difference is usually visible only visually in looking at phase-space, and is hard to quantify. As we show below, ordinal pattern analysis reveals a great deal about the underlying dynamics, including transitions hidden from Lyapunov exponent calculations.

We start our paper with an introduction to our model system, the classical and semi-classical Duffing oscillator, followed by a quick review of previous results from Poincar\'e sections and Lyapunov exponents. We then introduce the new techniques of ordinal pattern analysis and permutation entropy and present our results from a detailed analysis of the behavior of the classical and semi-classical systems as revealed by the new techniques, followed by a final conclusion.

\section{The Duffing Oscillator.}

The Duffing equation can be used to model various types of damped and driven oscillators. One physical manifestation is that of a clamped cantilever beam which has two magnets creating stable potential wells on each side of the beam with an unstable fixed point at the central location, and with this beam being periodically driven. Nano-electro-mechanical (NEMS) versions of such systems are now experimentally in a regime where semi-classical effects can be seen [6]. 

The starting point for the full analysis is the stochastic Schr\"odinger equation for a quantum trajectory of a system in a double-well potential with coupling to an environment that yielding both the environmental noise (fluctuation) and damping (dissipation) effects. The details are available in Ref.[3]. The analysis yields a scaling factor $\beta$, which is constructed to be dimensionless and arises from the system's parameters of mass $m$, length $l$, and natural frequency $\omega_0$ as $\beta^{2} = \frac{\hbar}{ml^{2}\omega_{0}}$. There is also a driving force of strength $g$ and frequency $\omega$, as well as a dissipative coupling to the environment $\Gamma$. It can be shown that $\beta\to 0$ yields the classical limiting equation. This limit can be understood as coming simply from increasing the system size (increasing $l$) for example. In this limit we get the equation 
\begin{equation}\label{eq:class}
\ddot{x} + 2\Gamma \dot{x} + \beta^2x^3 - x = \frac{g}{\beta}\cos(\omega t),
\end{equation}
which describes a classical particle with position $x$ and momentum $p=\dot{x}$. 
For intermediate length scales, we obtain a semi-classical approximation given by the five equations
\begin{equation}\label{eq:s1}
dx = p \, dt + 2 \, \sqrt[]{\Gamma}((\mu - \frac{1}{2})\, d\xi_R - R\,d\xi_I)
\end{equation}
\begin{equation}\label{eq:s2}
dp = (-\beta^2(x^3+3\mu x)+x-2\Gamma p + \frac{g}{\beta} \cos{(\omega t)}) \,dt + 2 \, \sqrt[]{\Gamma}(R\, d\xi_R-(\kappa-\frac{1}{2})\,d\xi_I)
\end{equation}
\begin{equation}\label{eq:s3}
\frac{d\mu}{dt} = 2R+2\Gamma(\mu-\mu^2-R^2+\frac{1}{4})
\end{equation}
\begin{equation}\label{eq:s4}
\frac{d\kappa}{dt} = 2R(-3 \beta^2x^3+1)+2\Gamma(-\kappa-\kappa^2-R^2+\frac{1}{4})
\end{equation}
\begin{equation}\label{eq:s5}
\frac{dR}{dt} = \mu(-3 \beta^2x^3+1)+\kappa-2\Gamma R(\mu+\kappa).
\end{equation}
As for the classical equations we have the centroid variables $x=\langle \hat{Q} \rangle$ and $p=\langle \hat{P} \rangle$. However, for the semi-classical system, the centroid variables couple to the spread variables $\mu=\sigma_{QQ}$, $\kappa=\sigma_{PP}$ and $R= \frac{1}{2}(\sigma_{PQ}+\sigma_{QP})$ where $\sigma_{AB} =\langle (\hat{A}^\dagger-\langle \hat{A} \rangle^*)(\hat{B}-\langle \hat{B} \rangle) \rangle$. The effect of environmental fluctuations acting on the quantum system is visible via $\xi_R$ and $\xi_I$ where $d\xi =d\xi_R + i d\xi_I$ is a normalized complex differential random variable satisfying $M(d\xi) =0; M (d\xi d\xi^*)=dt$ where $M(\cdot)$ denotes the mean over realizations.

Previous research [3] has compared the classical and semi-classical chaos in this system using the Lyapunov exponent $\lambda$. The bulk of the analysis focused on the situation with fixed $\omega=1,g=0.3$, while $\Gamma$ was varied to explore the different kinds of dynamics visible. The results can be summarized as follows: Classically, as $\Gamma$ initially increases away from zero the classical orbits remain simply periodic over the double well, with the driving compensating for the dissipation. For $\Gamma\approx 0.06$ there is a transition to chaos. At this stage, the dissipation is large enough such that orbits do not always cross the central potential barrier which can yield chaos since given two nearby orbits, one may cross the barrier while the other does not, leading to drastically different evolution. This persists until $\Gamma\approx 0.210$ after which the high damping restricts the dynamics entirely to one of the wells. While there is some single-well chaos for small window of $\Gamma > 0.210$ for sufficiently high damping, there is no complex behavior. Even when the system is mostly chaotic, there are many small windows of regularity, most prominently around $\Gamma\approx 0.110$ for example. Thus we see the system switching between periodic behavior and chaotic attractors as we scan $\Gamma$ although the differences between different kinds of periodic orbits or between the various chaotic attractors seen are not visible in the value of $\lambda$. The semi-classical Lyapunov exponents broadly follow the classical $\lambda$ as a function of $\Gamma$, except that the semi-classical system is {\em less parametrically sensitive} to environmental coupling. That is, $\lambda$ changes less for the semi-classical system as $\Gamma$ changes (compared to the classical behavior), and there are no abrupt transitions. Remarkably, there are also multiple regimes where the system has classical $\lambda < 0$ (no chaos) to semi-classical $\lambda >0$ (chaos) which -- contrary to standard intuition -- indicates that quantum effects can render a classical regular system chaotic. 

Given this intriguing result, it would be useful to find experimentally viable methods of investigating this behavior. In what follows we re-examine the system in this same regime with precisely this aim and using ordinal pattern analysis on the trajectories of the classical and semi-classical Duffing oscillator as we vary $\Gamma$. 
 

\section{Ordinal patterns analysis and permutation entropy.}

Ordinal pattern analysis [7-12] has been shown to be a powerful tool to analyze time series of events, and to explore time correlations and memory. It has been widely used to characterize symptoms in epileptic patients [13], or to classify cardiac biosignals [14], to study EEG signals[15,16], to distinguishing determinism from stochasticity in optical systems [17], to identify hidden time-scales [18], or to understand phase transitions [19]. This symbolic analysis method transforms the original time series into a sequence of patterns, based on the relative size of consecutive events, or inter-event intervals. We can then compute the probabilities of occurrence of the different patterns to study their hierarchy, entropy, or the existence of forbidden patterns.

The first step in ordinal patterns analysis is to take a time series of N events and transform it into a series of N-D time intervals (TI) of dimension D. Then, D consecutive TIs are compared and labeled depending on their relative size, i.e., for dimension D=2 only two patterns, also referred to as words, are possible: '01' and '10', for $TI(i)<TI(i+1)$, and $TI(i)>TI(i+1)$, respectively. For dimension D=3, six different words are possible: '012', for $T(i)<T(i+1)<T(i+2)$; '021', for $T(i)<T(i+2)<T(i+1)$; ... This transformation from events to words keeps the information about the time correlations in the sequence of events, and the short-time memory in the system, but it loses the information contained in the duration of the TIs. By computing the probabilities of appearance of the different words we can extract information about the dynamics of the system. For a purely stochastic process we expect all words to be equally probable ($1/6$ for words of dimension D=3), while a more spread out distribution of the probabilities would be related to a more deterministic dynamics.

\begin{figure}[tph]
   \centering
     \resizebox{0.6\columnwidth}{!}{\includegraphics{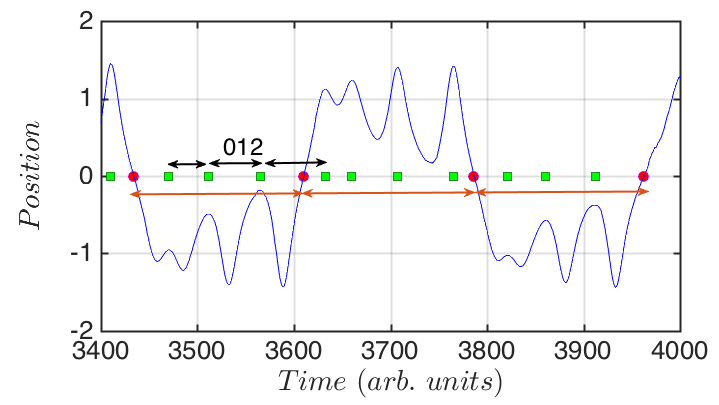}}
    \caption{Position of the oscillator versus time. The green squares indicate the detection of a peak, and the red dots indicate the crossing from one well to the other. Indicated with black arrows is depicted one word of dimension D=3 as example. The word is computed with inter-peak intervals. Because $TI(i)<TI(i+1)<TI(i+2)$, we designate the word as $012$. The orange arrows indicate three consecutive inter-crossing intervals. These time-intervals are equal, indicating certain regularity in the dynamics for crossing from one well to the other.
    \label{figure_words_example}}
\end{figure}

Figure \ref{figure_words_example} depicts a section of a time series of the double-well Duffing oscillator, showing how the events are transformed into words, or ordinal patterns. The position has been normalized so that zero refers to the mid-point between both wells, positive positions correspond to one of the wells and negative positions to the other well. Green squares indicate peak detections, while red circles indicate the crossing from one well to the other. By comparing three consecutive TIs we compute the words of dimension 3. Because, for the three consecutive peaks selected, $TI(i)<TI(i+1)<TI(i+2)$ the corresponding word is '012'. For the three consecutive crossings all TIs are equal. This reveals a periodic behavior of the system when jumping from one well to the other.

Once we know the probabilities of the ordinal patterns we can quantify the complexity of the dynamics through the permutation entropy (PE) [7]. This entropy is computed with the probabilities ($P_i$) of the different words of dimension D, and is normalized so that $0\leq PE \leq 1$, as
\begin{equation}
PE = -\sum_{i=1}^{D!} \frac{P_i~log(P_i)}{log(D!)}.
\end{equation}
PE has shown to be a helpful parameter in characterizing time series where more traditional parameters are not as conclusive, such as classifying cardiac time series, sleep stages from EEG recordings, or detecting laminar-to-turbulent transitions, among others [13,15,19].

\section{Lyapunov exponent and Poincar\'e sections.}

\begin{figure}[tph]
   \centering
     \resizebox{1.0\columnwidth}{!}{\includegraphics{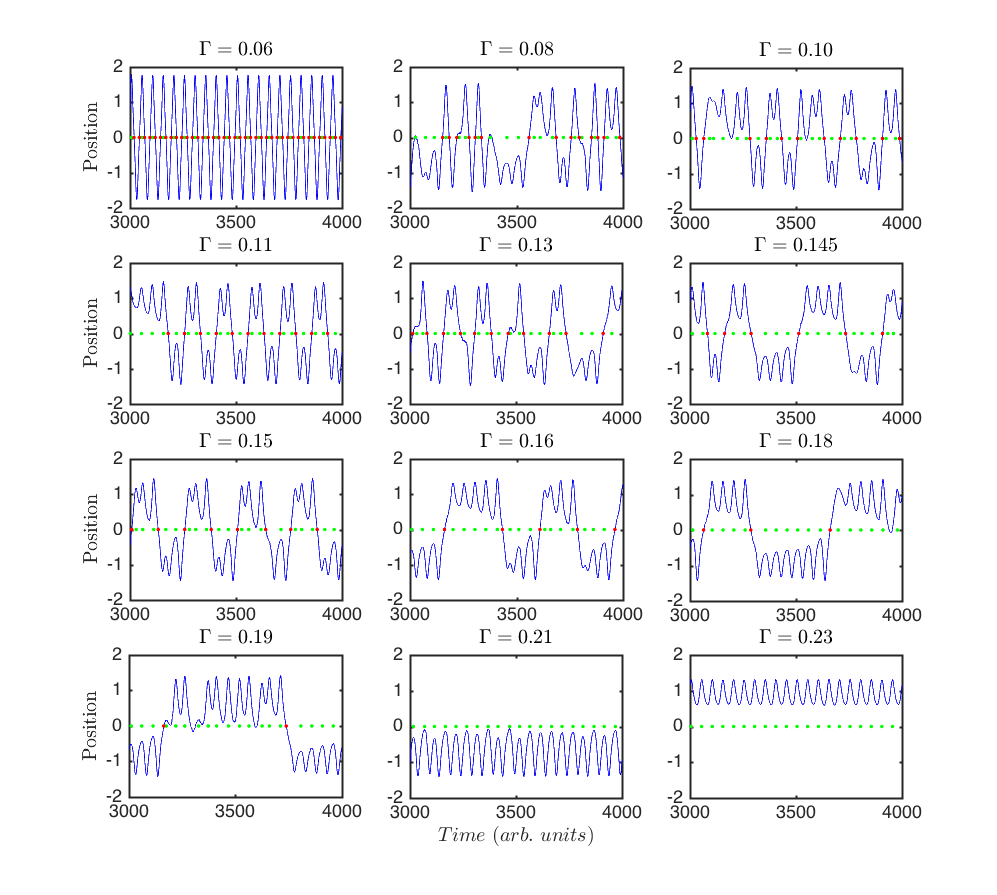}}
    \caption{Time series of the dynamical system for various values of the damping parameter, $\Gamma$. The green dots indicate the detection of a peak, and the red dots indicate the crossing from one well to the other.
    \label{figure_time_series}}
\end{figure}

Sections of the time series of the system can be seen in figure \ref{figure_time_series} for different values of the damping parameter, $\Gamma$. The plots show the position of the system versus time. It can be seen how the dynamics for $\Gamma = 0.06$ and $\Gamma = 0.23$ are periodic with period one, i.e., the system is driven by the periodic forcing. Between those extremes, the system presents a variety of behaviors, with different periodicities and transitions. Because the damping acts against the tendency of the system to jump from well to well, the greater the damping, the longer it takes the system to jump, and the longer it remains in the same well.

\begin{figure}[tph]
   \centering
     \resizebox{1.0\columnwidth}{!}{\includegraphics{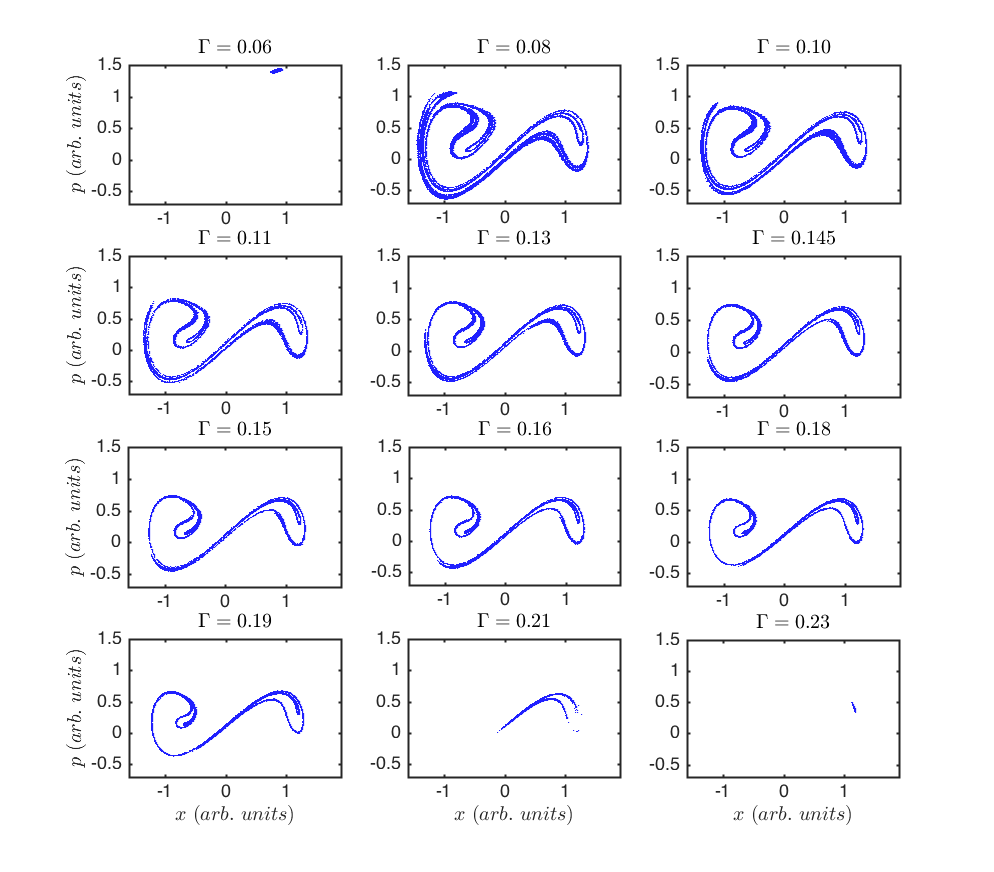}}
    \caption{Poincar\'e section for the various values of the damping parameter in figure \ref{figure_time_series}. Horizontal axis corresponds to position, $x$, and vertical axis to momentum, $p$. $\Gamma = 0.06$ and $\Gamma = 0.23$ reflect the periodic behavior while for $0.07 < \Gamma < 0.21$ the system is chaotic. For $\Gamma = 0.21$ this recurrence map shows how the system only moves in one of the two wells. Poincar\'e sections of the system do not indicate any transition in the chaotic regime. The attractor evolves smoothly as the control parameter, $\Gamma$, is changed. 
    \label{figure_poincare}}
\end{figure}

Figure \ref{figure_poincare} shows Poincar\'e sections for the same $\Gamma$ values used in figure \ref{figure_time_series}. Poincar\'e sections give information of the shape of the chaotic attractor of the system, we plot them computed with the frequency of the external driving. We recover the same results, i.e., for very low and very high damping the system is periodic, and this periodicity is the one of the external forcing; for other values of $\Gamma$ the system shows a non-periodic behavior, where the chaotic attractor seems to maintain shape while slowly shrinking, as $\Gamma$ is increased.

Figure \ref{figure_words_gamma}(a) displays the Lyapunov exponent ($\lambda$) of the double-well system versus the damping parameter, $\Gamma$. It shows that the system presents a regular behavior for low values of the damping (for $\Gamma<0.07$, $\lambda<0$), but it sharply changes to chaotic behavior, maintained until very high damping ($\Gamma>0.21$), where it becomes non-chaotic again. This is in agreement with figures \ref{figure_time_series} and \ref{figure_poincare}. We can see a direct relation between the regions where the Lyapunov exponent is negative and the periodic dynamics as visible in the time series and Poincar\'e sections.

\section{Results and discussion.}

\subsection{The Lyapunov exponent.}

Figures \ref{figure_words_gamma}a,\ref{figure_words_gamma}b show the Lyapunov exponent versus coupling $\Gamma$ for the classical (a), and semi-classical (b) Duffing oscillator. As described above, and in agreement with Figs. \ref{figure_time_series} and \ref{figure_poincare}, the system presents a regular behavior for $\Gamma<0.07$ and $0.212<\Gamma$. Between those limits, the classical Duffing oscillator presents a very diverse behavior. Chaos dominates but windows of periodicity appear. There are abrupt changes from chaos to periodicity and vice versa. As the system increases its coupling to the environment the behavior can be very different for very similar parameters. For the semi-classical Duffing oscillator although the abrupt windows of periodicity are not present, the exponent does resembles the classical one. However, in particular the system shows chaos for all values of coupling, $0.07<\Gamma<0.212$. This shows a smoothening of the graph in going from the classical regime to the semi-classical regime.

\subsection{Ordinal pattern analysis of the classical Duffing oscillator.}

For the classical Duffing oscillator, Lyapunov exponents before and after one of the periodicity windows are similar, but we can not tell if the dynamics of the system are similar. In order to study the complex dynamics of the system in more depth we compute the ordinal patterns of dimension 3. Through the ordinal patterns we can explore the system in more detail, as we have now six different parameters to do it. Also, these ordinal patterns can unveil temporal correlations in the system for different $\Gamma$ regimes.

\begin{figure}[tph]
   \centering
     \resizebox{0.9\columnwidth}{!}{\includegraphics{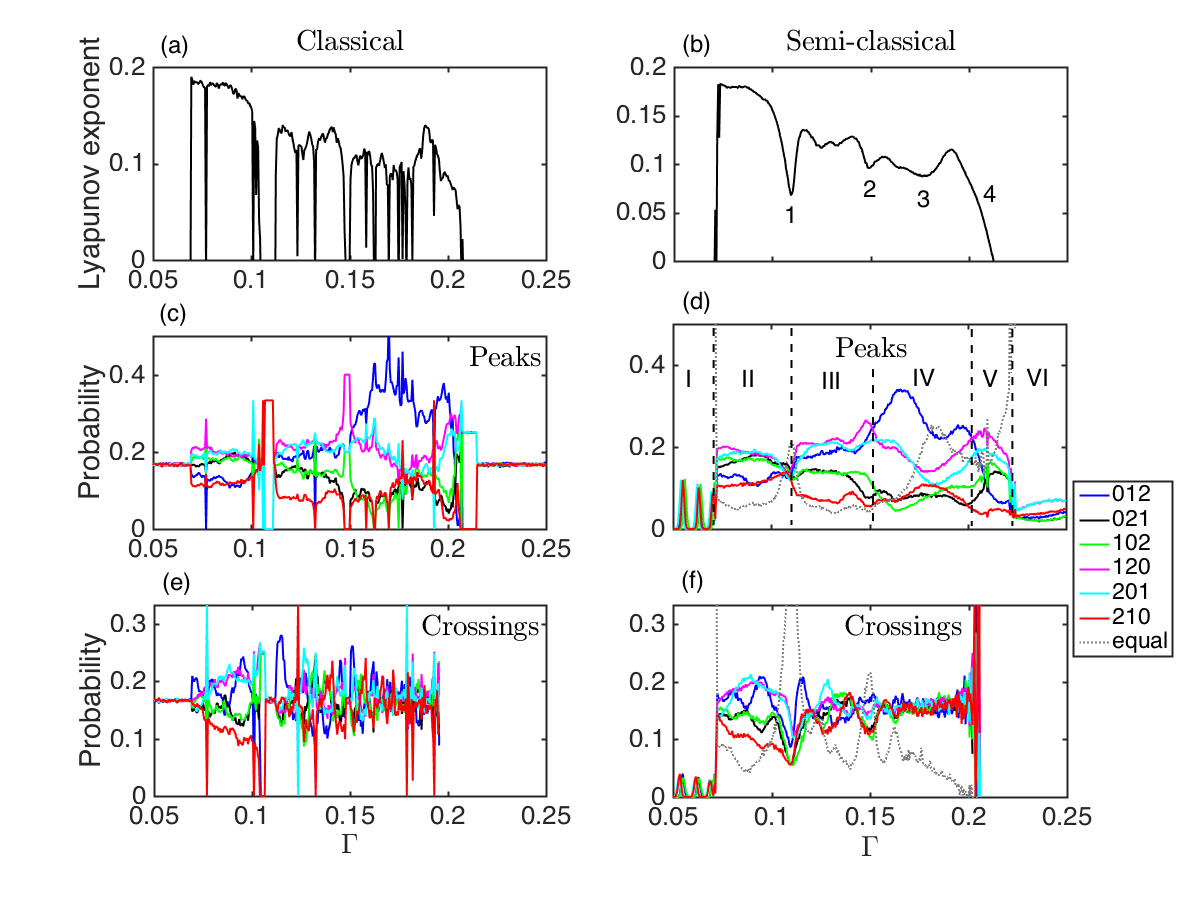}}
    \caption{(\textbf{a}) Lyapunov exponent versus the damping parameter, $\Gamma$. (\textbf{b}) Words probabilities of dimension 3 considering the peaks of the time series as events. Six regions can be differentiated from the distinct hierarchies of the words. (\textbf{c}) Words probabilities of the dynamics considering as events the crossings from one potential well to the other potential well. The gray dotted line corresponds to the words that contain one, two or three equal consecutive time intervals.
    \label{figure_words_gamma}}
\end{figure}

To compute the words we consider the timing of events in the time series (see figure \ref{figure_words_example} for event detection and words computing). These events can be (i) the peaks of the time series (figure \ref{figure_words_gamma}c,\ref{figure_words_gamma}d, and green squares in figures \ref{figure_words_example} and \ref{figure_time_series}), or (ii) the times of the crossings from one well to the other (figures \ref{figure_words_gamma}e,\ref{figure_words_gamma}f, and red dots in figures \ref{figure_words_example} and \ref{figure_time_series}). This analysis shall reveal temporal correlations in the global dynamics (peaks), or in the inter-well dynamics (crossings). Figures \ref{figure_words_gamma}d, and \ref{figure_words_gamma}f include a seventh word (dotted line), computed with those inter-event intervals where there are two or three equal consecutive intervals ($TI(i-1)=TI(i)$, see figure \ref{figure_words_example}). This periodic seventh word dominates when the system has a periodic-type behavior.

By comparing Figs. \ref{figure_words_gamma}c,\ref{figure_words_gamma}d we can see that the dynamics of the semi-classical Duffing oscillator retains a memory of the classical system, as also reflected in the Lyapunov exponent. For the values of $\Gamma$ where there is chaos in the classical case, the hierarchies of the words probabilities are the same in the semi-classical. For the windows of periodicity in the classical system we can see this periodicity reflected in the appearance of forbidden words, and in the values of the probabilities of the allowed words, i.e., for $\Gamma = 0.11$ the probabilities are 1/3 for words '021', '102' and '210', while '012', '120', and '201' are forbidden; for the window of periodicity around $\Gamma = 0.148$ the probabilities are 1/4 for '120', 1/5 for '012', '102', and '201', while '021' and '210' are forbidden.

Investigating the word hierarchies of the classical system we see that they are not the same before and after a window of periodicity. This change of hierarchy shows a different chaotic dynamics for those parameters. As we increase the control parameter the system abandons the chaotic behavior to show periodicity, and when the systems returns to chaos it performs a different type of dynamics. The ordinal patterns probabilities indicate that the system does not explore the same region of the attractor before and after the periodicity window, even though the parameters of the system and the Lyapunov exponent are very similar. This changes in behavior is not evident if we only use the Lyapunov exponent.

All the sharp drop-outs and spikes in the words probabilities correspond to the $\Gamma$ values where the system leaves chaos, as seen with Lyapunov exponent. Thus we do not need to calculate Lyapunov exponents to distinguish when the system is chaotic or periodic, and further this ordinal pattern analysis also reveals when, for the same Lyapunov exponent, the system performs a different dynamics.

\subsection{Ordinal pattern analysis of the semi-classical Duffing oscillator.}

The semi-classical system is more interesting, as there are no windows of periodicity to indicate a drastic change in behavior, but we can also see the change in the hierarchies of the words. We can determine six different regions depending on the type of dynamics described by the words probabilities (see Fig. \ref{figure_words_gamma}d).

Regions I ($\Gamma < 0.07$), and VI ($0.212< \Gamma$), correspond to non-chaotic dynamics, in agreement with the time series (figure \ref{figure_time_series}), Poincar\'e sections (figure \ref{figure_poincare}), and Lyapunov exponent (figure \ref{figure_words_gamma}a, $\lambda<0$). This is also indicated by the presence of the seventh word, which contains equal consecutive inter-event intervals. The abundance of this seventh word is a signature of periodicity, and its probability is one, or almost one, for regions I and VI in the peaks, and for region I for the crossings (there are no crossing events in region VI, where damping dominates such that the system remains in the same well and can not cross the central barrier). In regions I and VI the forcing is driving the system.

\begin{figure}[tph]
   \centering
     \resizebox{0.6\columnwidth}{!}{\includegraphics{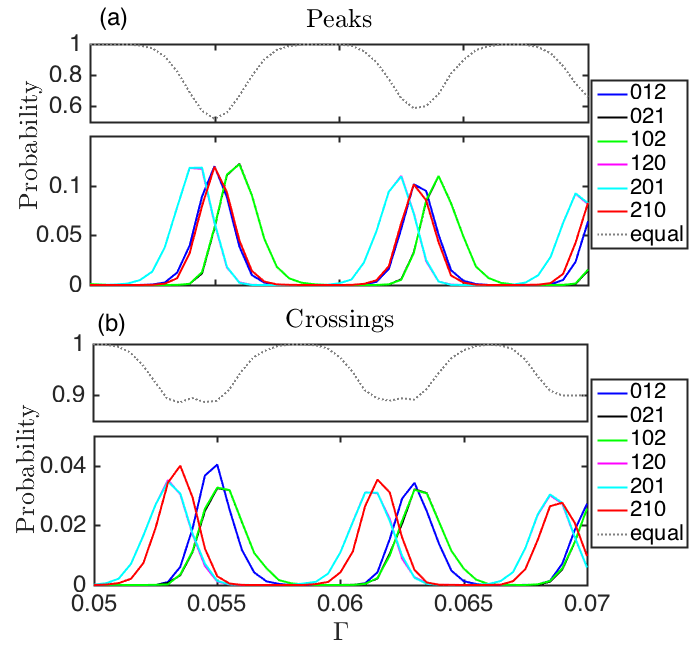}}
    \caption{Words probabilities computed with the peaks (a) and with the crossings (b) for region I. The dotted line corresponds to the word with consecutive equal time intervals. The probability for the periodic word is much higher than for the rest of the words.
    \label{figure_words_region_I}}
\end{figure}

In region I two differentiated regimes can be seen. For some values of $\Gamma$ only the seventh word is present (completely periodic dynamics), while for other $\Gamma$ values some other words have non zero probability, although small compared to the seventh word (see figure \ref{figure_words_region_I}).

For region II ($0.07 < \Gamma < 0.11$) in figure \ref{figure_words_gamma}(b), the words probabilities present two differentiated groups, with the words '012' and '210' being less probable than average, and the rest  more probable than the average. The less probable words are those that keep a persistence, as they are formed with consecutive increasing ('012') and consecutive decreasing ('210') time intervals. In this region, therefore, the most probable patterns are those that present up-down, time-oscillatory behavior. This is a signature of the proper frequency of the system competing with, but being dominated by, the external forcing. This hierarchy of the words can also be found in a chaotic optical system. A diode laser with optical feedback and external periodic forcing, in the regime where the external forcing is the dominant term, also shows this hierarchy of the words (see figure 4 of ref. [17]). This is in agreement with the fact that in region II of Fig. \ref{figure_words_gamma} forcing is dominant over damping.

Region III ($0.11 < \Gamma < 0.15$) shows a different dynamic regime. The hierarchy of the words is different from the one in region II. The words probabilities are more spread out. This is related with a more deterministic, less stochastic, behavior. As the damping is increased, the system spends more time in each well before jumping to the other well. The in-well dynamics becomes more relevant with respect to the inter-well dynamics, as the damping increases. The transition between region II and region III is sharp, at $\Gamma=0.11$, and it is characterized by an increase of the periodicity of the system, despite being still in a chaotic regime ($\lambda > 0$). The seventh word is the most relevant in the crossings and it increases considerably in the global, peaks dynamics (to 0.2). Looking at the behavior of the Lyapunov exponent (figure \ref{figure_words_gamma}b), while still positive, we see a dip right where the words indicate a transition in the dynamical behavior.

For region IV ($0.151<\Gamma < 0.21$) the word '012', consecutive more spaced time intervals, is more likely to happen than the others, while the word '210', consecutive less spaced time intervals, remains among the less probable words. In this region, for the largest values of $\Gamma$, the system begins to lose information of the crossings; it spends increasing amounts of time, as $\Gamma$ increases, in one well and it rarely jumps to the other well. Therefore, the hierarchy of the words in the peaks has a negligible influence from the second well. There is also an increase in the underlying periodicity of the dynamics (seventh word) as the system reduces its jumps from well to well.

Region V corresponds to the transition from chaotic to regular behavior ($\lambda$ goes from positive to negative). All the dynamics is due to a single well, as there are no crossings in this regime (see figure \ref{figure_words_gamma}f). The damping does not allow the system to jump between wells. The words distribution in regions V and VI is due only to the topology of the attractor of one of the wells. Here the word '012' in the peaks dynamics, that was the most probable in region IV, becomes less probable than average, together with '210'. This means a loss of persistence in the dynamics, just as in region II. Indeed, the hierarchy of the words is the same as in region II, besides the fact that the seventh word becomes increasingly relevant as the system transitions to the non-chaotic behavior. The fact that regions II and V have the same hierarchies of the words can be interpreted as similar behavior of the system. In region II the damping is small and the barrier between wells has a small effect compared with the overall energy of the system, i.e., the system behaves as in a single well. In region V, damping is strong and the system is bounded to a single well from which it can not escape.
Finally,as the damping becomes more and more intense the system becomes more periodic, until it is finally free of chaos. Region VI corresponds to the non-chaotic dynamics of one well with external forcing and large damping.

Because the seventh word contains signatures of periodicity, it is clear that when its probability is high the dynamics has a relevant regular component. This regular signature can also appear underlying the chaotic region. Indeed it happens for the crossings at the transitions between regions I-II, II-III, and III-IV. The peaks also capture this increase of word seven for the transitions I-II, II-III and V-VI.

Looking at the Lyapunov exponent right at these transitions unveiled by the ordinal patterns, we can interpret the dips in the exponent as changes of the underlying chaotic dynamics. Dips 1, 2 and 4 in figure \ref{figure_words_gamma}b correspond to transitions II-III, III-IV, and V-VI, the latter being the transition from chaos to non-chaos. Also, dip 3 is reflected in the increase of the seventh word of the peaks, which reflects a more regular behavior.

\begin{figure}[tph]
   \centering
     \resizebox{0.6\columnwidth}{!}{\includegraphics{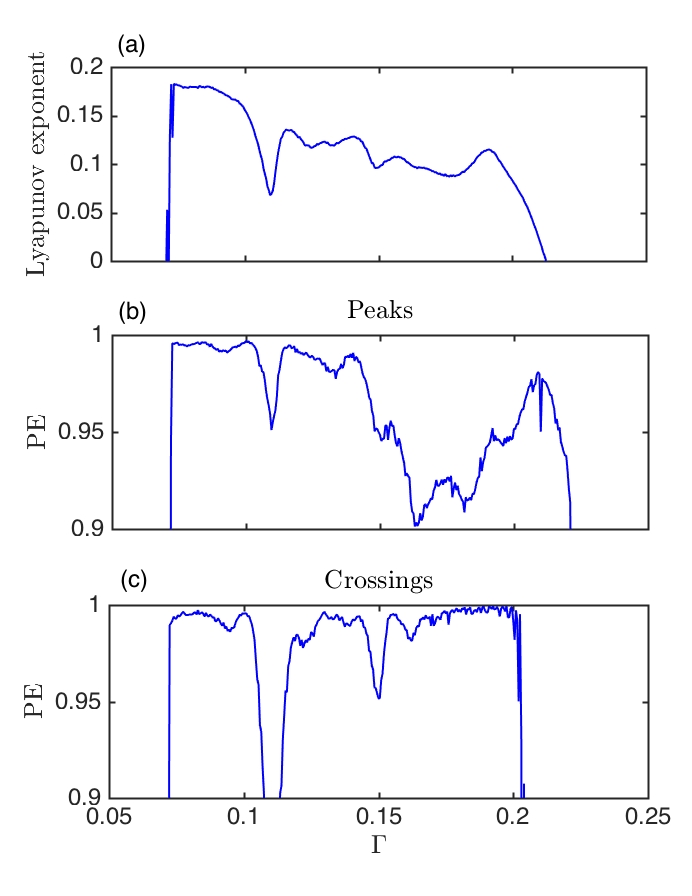}}
    \caption{(\textbf{a}) Lyapunov exponent versus the damping parameter, $\Gamma$. (\textbf{b}) Permutation entropy (PE) computed with words of dimension 3 with the peaks. (\textbf{c}) PE computed with words of dimension 3 with the crossings.
    \label{figure_PE}}
\end{figure}

\subsection{Permutation entropy and relation with Lyapunov exponent.}

Figure \ref{figure_PE} shows the Lyapunov exponent (a) and the PE computed with words of dimension 3 for the semi-classical system (the seven different words used in figure \ref{figure_words_gamma}) for the peaks (b) and the crossings (c). PE captures qualitatively the variations of the Lyapunov exponent for low to intermediate $\Gamma$. For intermediate to high $\Gamma$ the crossings can not capture any changes in the dynamics, as there are not enough crossings.  But the permutation entropy of the peaks do capture the complex transitions in dynamics reflected in the words probabilities in this regime. Thus PE not only complements Lyapunov exponent calculation, but it also goes beyond in exposing the complexity of a dynamical system.

\section{Conclusions.}

We have studied the complex dynamics of the classical and semi-classical double-well Duffing oscillator, using the techniques of ordinal patterns analysis and permutation entropy. We have shown that these techniques quantitatively indicate the same transitions between chaos and regular behavior as the better-known technique of Lyapunov exponent, as a function of $\Gamma$, the coupling to the environment, as well as as a function of $\beta$, a measure of the system size. However, these techniques deployed here only need a single (even if noisy) time-series from a single observable, which is significantly simpler to access in experimental situations.

We have also seen that the system performs different dynamics as we vary the damping of the system. These different dynamics are not revealed by the Lyapunov exponent, but the ordinal patterns clearly detect the transitions in the dynamics, indicated by different hierarchies of the patterns in the different regimes revealed.

We have studied the dynamics by looking at two different types of events: the peaks of the time series, that capture the global dynamics of the system; and the crossings of the system from one well to the other well, that focuses in the inter-well dynamics. The hierarchies of the words are different in each set of events (peaks or crossings), but the regions and the transitions are revealed with both.

While the classical Duffing oscillator presents windows of periodicity, and the semi-classical displays chaos for the most of the parameter region analyzed, the semi-classical system remembers the classical through the Lyapunov exponent and the hierarchies of words. But the ordinal analysis can detect transitions and distinguish between dynamics, even when the system does not transition in and out of regular dynamics, and even when the Lyapunov exponents are only slightly altered.

By computing the permutation entropy of dimension 3 for peaks and crossings we have seen a relationship between PE and Lyapunov exponent that lets us conclude that the dips in the latter indicate transitions in the underlying chaotic dynamics. These transitions and changes in the dynamics can not be detected by exploring the time series or the Poincar\'e sections.

All of the above shows that the techniques of ordinal patterns analysis and permutation entropy are a computationally simple and reliable tool to explore the complexity of a system's dynamics.

\section{Acknowledgments}

M. T. gratefully acknowledges financial support from Carleton College via internal funds. \\

\section{References}

[1] Spiller, T. \& Ralph, J. The emergence of chaos in an open quantum system. {\it{Physics Letters A}}, { \textbf{194}}, 235-240 (1994).

[2] Brun, T., Percival, I. \& Schack, R. Quantum chaos in open systems: a quantum state diffusion analysis.\emph{J. Phys. A-Math. Gen.} { \bf  29,} 2077-2090 (1996).

[3] Bibek Pokharel, Peter Duggins, Moses Misplon, Walter Lynn, Kevin Hallman, Dustin Anderson, Arie Kapulkin, Arjendu K. Pattanayak, Dynamical complexity in the quantum to classical transition, arXiv:1604.02743v1 (2016).

[4] Boffetta, G., Cencini, M., Falcioni, M. \& Vulpiani, A. Predictability: a way to characterize complexity. \emph{Phys. Rep.} { \bf 356,} 367-474 (2002).

[5] Ralph, J., Jacobs, K. \& Everitt, M. Observing quantum chaos with noisy measurements and highly mixed states. \emph{Phys. Rev. A} { \bf 95,} (2017).

[6] Li, Q., Kapulkin, A., Anderson, D., Tan, S., and Pattanayak, A. Experimental signatures of the quantum-classical
transition in a nanomechanical oscillator modeled as a damped-driven double-well problem. \emph{Phys. Scripta}, { \bf T151}, 014055 (2012).

[7] Bandt and Pompe, Quantum interference between two single photons emitted by independently trapped atoms. {\it Phys. Rev. Lett.} {\bf{440}}, 779-782 (2012).

[8] J. M. Amig\'o, K. Keller, and J. Kurths, Recent progress in symbolic dynamics and
permutation complexity: ten years of permutation entropy. {\it Eur. Phys. J. Spec. Top.} {\bf{222}}, 2 (2013).

[9] Massimiliano Zanin, Luciano Zunino, Osvaldo A. Rosso, and David Papo, Permutation Entropy and Its Main Biomedical and Econophysics Applications: A Review.  {\it Entropy} {\bf{14}}, 1553-1577 (2012)

[10] M. Riedl, A. M\"uller, and N. Wessel, Practical considerations of permutation entropy.  {\it Eur. Phys. J. Special Topics} {\bf{222}}, 249-262 (2013)

[11] Douglas J. Littlele and Deb M. Kane, Variance of permutation entropy and the influence of ordinal pattern selection. {\it Phys. Rev. E.} {\bf{95}}, 052116 (2017).

[12] Haroldo V. Ribeiro, Max Jauregui, Luciano Zunino, and Ervin K. Lenzi, Characterizing time series via complexity-entropy curves. {\it Phys. Rev. E.} {\bf{95}}, 062106 (2017).

[13] M. Zanin, M. Zunino, O. A. Rosso, and D. Papo, Permutation entropy and its main biomedical and econophysics applications: a review. {\it Entropy} {\bf{14}}, 1553 (2012).

[14] U. Parlitz, S. Berg, S. Luther, A. Schirdewan, J. Kurths, N. Wessel. Classifying cardiac
biosignals using ordinal pattern statistics and symbolic dynamics. {\it Comput. Biol. Med.} {\bf 42},
319?327 (2012).

[15] C. Bandt, A New Kind of Permutation Entropy Used to Classify Sleep Stages from Invisible EEG Microstructure. {\it Entropy} {\bf{19}}, 197 (2017).

[16] Karsten Keller, Teresa Mangold, Inga Stolz, and Jenna Werner, Permutation Entropy: New Ideas and Challenges. {\it Entropy} {\bf{19}}, 134 (2017).

[17] A. Aragoneses, S. Perrone, T. Sorrentino, M. C. Torrent, and C. Masoller, Unveiling the complex organization of recurrent patterns in spiking dynamical systems. {\it Sci. Rep} {\bf{4}}, 4696 (2014).

[18] L. Zunino, M. C. Soriano, I. Fischer, O. A. Rosso, and C. R. Mirasso, Permutation-information-theory approach to unveil delay dynamics from time-series analysis. {\it Phys. Rev. E} {\bf{82}}, 046212 (2010).

[19] A. Aragoneses, L. Carpi, N. Tarasov, D. V. Churkin, M. C. Torrent, C. Masoller, and S. K. Turitsyn, Unveiling temporal correlations characteristic to phase transition in the intensity of a fibre laser radiation. {\it Phys. Rev. Lett.} {\bf{116}}, 033902 (2016).

[20] A. B. Neiman and D. F. Russell, Models of stochastic biperiodic oscillations and
extended serial correlations in electroreceptors of paddlefish. {\it Phys. Rev. E.} {\bf{71}}, 061925 (2005).

[21] L. Carpi and C. Masoller, Persistence and Stochastic Periodicity in the Intensity Dynamics of a Fiber Laser During the Transition to Optical Turbulence (under revision).


\end{document}